\documentclass[twocolumn,latterpaper,superscriptaddress,groupedaddress, 10pt]{revtex4}  

\usepackage{graphicx}  
\usepackage{color}
\usepackage{dcolumn}   
\usepackage{bm}        
\usepackage{amssymb}   
\usepackage{amsmath,textcomp}
\newcommand{\agr}[1]{{\color{black}#1}}

\usepackage{natbib}
\usepackage[utf8]{inputenc}
\begin{document}

\title{A Stable Optical Trap from a Single Optical Field Utilizing Birefringence}

\author{Robinjeet Singh}
   \email{roubein@gmail.com}
   \affiliation{Department of Physics \& Astronomy, Louisiana State University, Baton Rouge, LA, 70808}
	\author{Garrett D. Cole}
	\affiliation{Vienna Center for Quantum Science and Technology (VCQ), Faculty of Physics, University of Vienna, A-1090 Vienna, Austria}
	\affiliation{Crystalline Mirror Solutions LLC and GmbH, Santa Barbara, CA, and Vienna, Austria}
	
	\author{Jonathan Cripe}
   \affiliation{Department of Physics \& Astronomy, Louisiana State University, Baton Rouge, LA, 70808}
	
	\author{Thomas Corbitt}
	\affiliation{Department of Physics \& Astronomy, Louisiana State University, Baton Rouge, LA, 70808}
	
	\date{\today}
\date{\today}

\begin{abstract}
We report a stable double optical spring effect in an optical cavity pumped with a single optical field that arises as a result of birefringence. One end of the cavity is formed by a multilayer Al$_{0.92}$Ga$_{0.08}$As/GaAs stack supported by a microfabricated cantilever, with a natural mode frequency of $274$ Hz. The optical spring shifts the resonance to $21$ kHz, corresponding to a suppression of low frequency vibrations by a factor of about $5,000$. The stable nature of the optical trap allows the cavity to be operated without any external feedback and with only a single optical field incident. 
\end{abstract}

\maketitle

Cavity opto-mechanics, the interaction of radiation pressure with movable optical elements, is an important field of study in gravitational-wave (GW) interferometers \cite{1-1,2-2,9} and in probing quantum mechanics with macroscopic systems \cite{1,2,3,4,5,6,10}. It is well established that in an opto-mechanical cavity, the radiation pressure due to the circulating field can act as a (anti-)restoring and (anti-)damping force, depending on whether the cavity is red or blue detuned \cite{11,12,13,14}. The (anti-)restoring force is generated by the position dependent intracavity power and radiation pressure, while the (anti-)damping force is due to the finite response time of the cavity to changes of the mirror position.

If the cavity length is adjusted so that its resonant frequency is less than the laser frequency (blue detuned), the radiation pressure gives rise to a positive restoring force and an anti-damping force. Likewise, when red detuned, anti-restoring and positive damping forces are generated. For systems in which the optical forces dominate their mechanical counterparts, this leads to instability from either an anti-restoring or anti-damping force. The relative signs of the restoring and damping may be modified when operated in the resolved-sideband regime \cite{6}, but here we focus on the regime in which the optical spring is much stronger than the mechanical stiffness, and the resulting optical spring resonance is at a lower frequency than the cavity linewidth. The optical spring formed by a restoring force  has a profound effect in systems with soft mechanical suspensions and can be used to enhance the sensitivity of detection by amplifying the mirror's motion. The strong anti-damping force can dominate the mechanical damping in this scenario giving rise to dynamic instabilities \cite{2-2,15,16} and is usually stabilized by actively controlling the optical response of the cavity through feedback loops \cite{2-2,15}. 

In 2007, Corbitt et. al. introduced a dual carrier stable optical trap, in which a damping force due to a red detuned sub-carrier field cancels out the anti-damping force due to the blue detuned carrier field \cite{17}. That approach eliminated the need for electronic feedback, but required using two distinct optical fields incident on the cavity. Recently, a new approach that exploits the bolometric backaction due to the photothermal effect was proposed by Kelley et. al. \cite{18}. This approach produces a damping force by exploiting the thermal expansion of the mirrors from absorption of the intracavity optical field. Though stable, such optical absorption introduces excess vacuum fluctuations and deteriorates the sensitivity of the device.


In this paper we introduce a new scheme to achieve a stable optical trap by exploiting the birefringence inherent to the mirrors, without relying on absorption or multiple carrier fields. We inject a single field with linear polarization into the cavity. The cavity consists of a 0.5 inch input mirror and a microfabricated mirror supported on a cantilever as the end mirror. The microresonator is fabricated from a stack of crystalline Al$_{0.92}$Ga$_{0.08}$As/GaAs layers and is inherently birefringent, resulting in differing resonance conditions for the orthogonal polarizations. \agr{The observed birefringence is in part a consequence of the finite lattice mismatch in the high and low index layers of the epitaxially grown distributed Bragg reflector structure of the microresonator \cite{cole13, cole14}}. The fabrication of the microresonator is described in the Supplemental Material \cite{SUPP}. 

The two polarization components of the input field undergo a relative phase shift as a function of the birefringence. This phase shift allows the two polarization components to operate at different cavity detunings, which gives rise to the stable double optical spring. We note that the phase shifted polarizations behave as if there were two input fields. We will refer to these orthogonal polarization components as the carrier (C) and the subcarrier (SC) polarizations, for convenience.

The schematic shown in Fig 1 describes the experiment performed to demonstrate our scheme. Initially the intensity of the laser field from the Nd:YAG laser is modulated by an amplitude modulator through a servo-controlled feedback signal from the transmitted cavity output field. The feedback provides a damping force to stabilize the optical spring while it is in the unstable region, and it only acts in a narrow frequency band around the optical spring resonance. The optical spring suppresses the cavity fluctuations below the optical spring resonance, up to a maximum factor of about $5,000$ at low frequencies, as determined by the ratio of the optical spring constant to the mechanical spring constant. That reduction stabilizes the cavity, and allows for long term operation without feedback at low frequencies. The polarization angle of the input field is set using a combination of two half wave plates and a polarizing beam splitter, such that the power in the C polarization is about 22 times the power in the SC polarization. \agr{The input power coupled to the cavity in C and SC polarizations is about 40.1 mW and 1.9 mW, respectively. }  

	\begin{figure}[t!]
\includegraphics[width=\columnwidth]{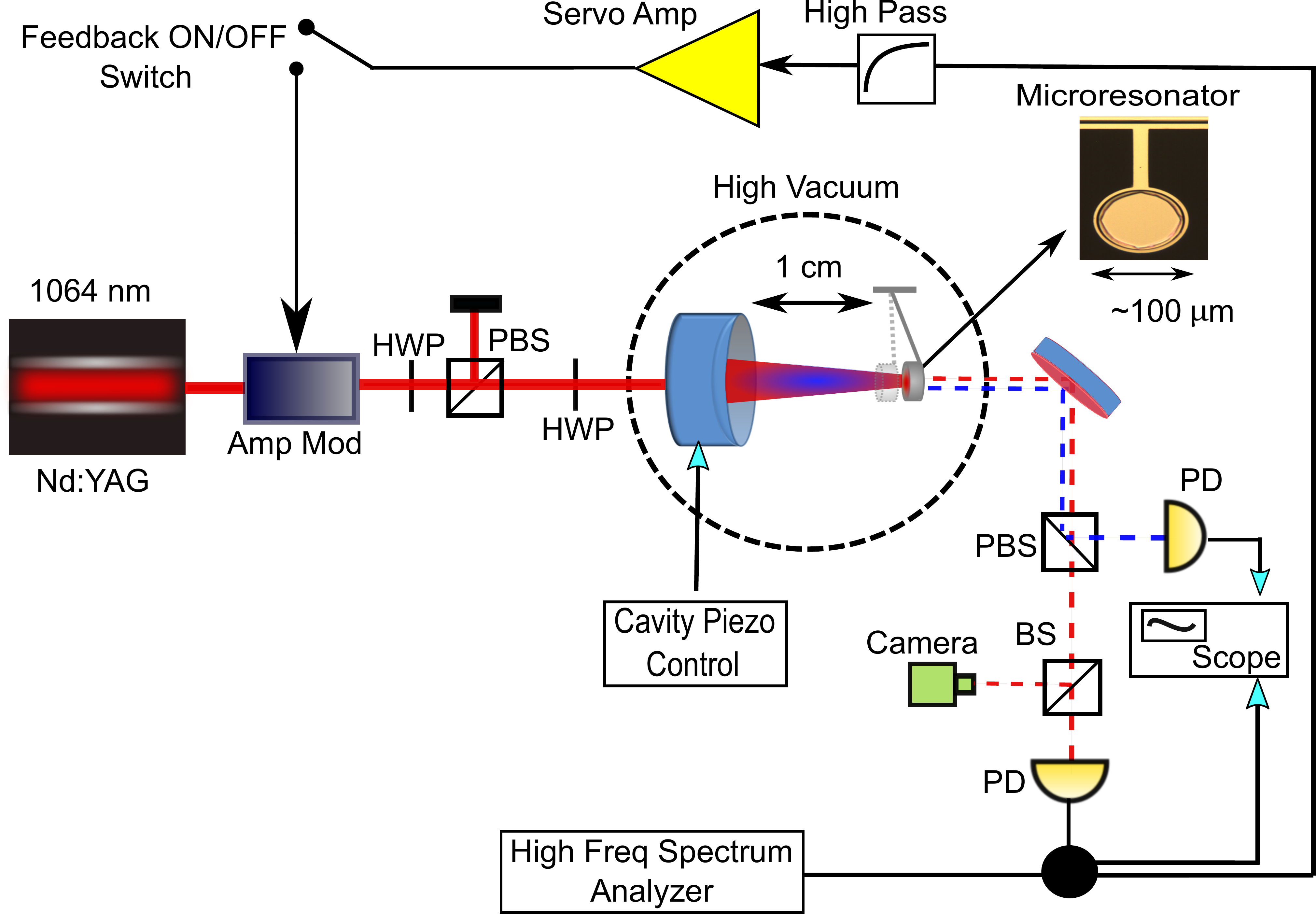}
\caption{Experimental setup: A 1064 nm Nd:YAG laser outputs 500 mW of near infrared light. The intensity of the input laser field is controlled by the amplitude modulator (Amp Mod) through a high pass feedback control loop. The first half wave plate (HWP) and a polarization beam splitter (PBS) sets the total coupled power of the input laser field to about \agr{42 mW} and the second HWP controls the power ratio between the carrier (C) and subcarrier (SC) polarization components of the input field to about 22:1. The cavity is located inside a vacuum tank and consists of a 0.5-inch diameter input mirror and the 100 \textmu m diameter microresonator (inset). The transmitted signals from the carrier (red) and the sub-carrier (blue) components are separated by a PBS. A 90:10 beam splitter (BS) splits the carrier transmission for signal detection by a photodetector and for qualitative detection by a camera. The carrier photodetector signal is used for signal analysis and as an error signal for the feedback control.}    
 \label{SETUP}
\end{figure}

The in-vacuum  cavity is one centimeter long and consists of an input mirror that has a radius of curvature of one centimeter. The input mirror is mounted on a piezoelectric device to allow for fine tuning of the cavity length. The optical field is focused on a microresonator that is about 100 \textmu m in diameter, and about 400 nanograms in mass. The microresonator has a natural mechanical frequency of $\Omega_m=2\pi\times274$ Hz with a mechanical quality factor $Q_m\approx2\times10^4$. The birefrengence induced frequency shift of the resonance condition between the two polarizations in our experiment is measured to be about 7.4 times the cavity linewidth (HWHM) of $\gamma \approx 2\pi\times254$ kHz. 
	
	The transmitted field from the end mirror is used to qualitatively analyze the cavity modes, determine the cavity noise spectrum, and to generate a feedback error signal for the initial control of the cavity. The C and the SC components of the transmitted fields are separated using a polarizing beam splitter, and the amplitude of the SC transmission is measured by a photodetector. The transmitted C polarization is further split by a 90:10 beam splitter for which 10\% of the signal is detected by a CCD camera in order to realize a qualitative analysis of the cavity modes. The rest of the C transmission is detected by a photodetector and is used both for the initial feedback control and the signal analysis of the cavity features. The electronic feedback control to the intensity of the input field is turned off once the self-stable regime is reached. 
	
	The power inside the cavity and the resulting radiation pressure on the microresonator test mass depends on the resonance condition of the cavity. For a large cavity linewidth, we take the frequency of motion $\Omega \ll\gamma$, such that the associated spring constant is given by \cite{18}

\begin{equation}
\mathrm{K_{os}} = \dfrac{16\pi{P_{in}}{T_1}\sqrt{R_1R_2^3}}{c\lambda_{o}(1-\sqrt{R_1R_2})^3}\dfrac{\delta_{\gamma}}{\left(1+\delta_{\gamma}^2\right)^2}
\label{1}
\end{equation}

where $P_{in}$ is the input power of the laser field. $T_i$ and $R_i$ are the transmittance and the reflectance of the input mirror $(i=1)$ and the end test mass $(i=2)$, $\delta_{\gamma}=\delta/\gamma$ is the field detuning in terms of the cavity linewidth, $\lambda_o$ is the center wavelength of the input laser field, and c is the velocity of light.  

In addition, the detuned cavity has a finite response time on the scale of $\gamma^{-1}$ and hence the intracavity power buildup lags the mirror motion. This lag in effect leads to a viscous damping force with a damping coefficient given by \cite{12,15}, again under the assumption that $\Omega\ll \gamma$:

\begin{equation}
\Gamma = \dfrac{-2\mathrm{K_{os}}}{M\gamma{[1+\delta_{\gamma}^2]}}
\end{equation}	
where M is the reduced mass of the two cavity mirrors. Compared with the fixed mirror the microresonator has a negligible mass and hence the reduced mass is simply equal to the mass of the cantilever.

For the optomechanical dynamics to be stable, a positive spring constant ($K>0$) and a positive damping coefficient ($\Gamma>0$) are required.  But as is evident from the dependence of K and $\Gamma$ on the sign of $\delta$ (Eq 1 and 2), a positive (restoring) spring constant implies instabilities due to negative damping force, under the assumption that $\Omega \ll \gamma$. This instability due to negative damping usually requires feedback control.

In our experiment, the system is stabilized by adjusting the detuning of the C and SC components of the intracavity field such that the blue detuned C polarization component creates a large restoring force and only small anti-damping force, while the red detuned SC polarization creates a small anti-restoring force and a large damping force. The reflectivities of the mirrors are the same for both polarizations in this system, as determined by optical ringdown measurements. At detunings of $\delta_C \approx 5.3\gamma$ and $\delta_{SC} \approx -2.1\gamma$, the intracavity carrier and  subcarrier polarizations component fields interact with the mechanical system resulting in $\mathrm{K_{tot}\Rightarrow \agr{K^C_{os} + K^{SC}_{os}>0}}$ and $\Gamma_{tot}\Rightarrow \Gamma_C + \Gamma_{SC}>0$.

\begin{figure}[t!]
\begin{centering}
\includegraphics[width=\columnwidth]{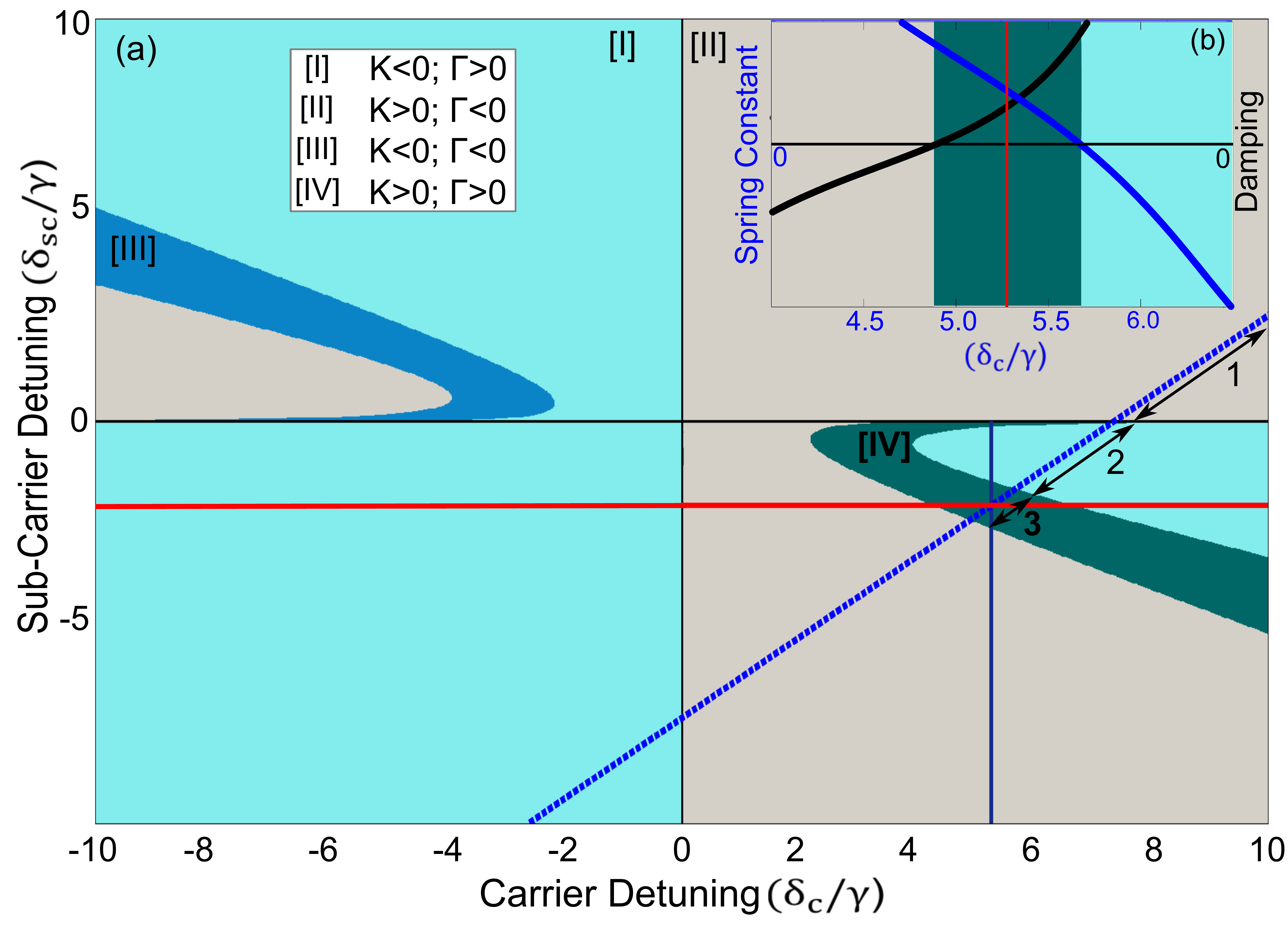}
\caption{Graphical representation for the total optical rigidity as a function of detunings of the carrier C and subcarrier SC at a fixed input power ratio of 22:1, respectively. The shaded regions [I], [II], [III], and [IV] respectively correspond to statically unstable region with K\textless0 and $\Gamma$\textgreater0, dynamically unstable region with K\textgreater0 and $\Gamma$\textless0, anti-stable region with K\textless0 and $\Gamma$\textless0, and stable region with K\textgreater0 and $\Gamma$\textgreater0. The dotted blue line represents the trajectory of the C over the cavity resonance and agrees with both the calculated and the experimentally measured data. Regions 1, 2, and 3 on the trajectory of C are in direct correspondence with the real time sweep data as shown in Fig 3.  A stable optical trap is achieved at $\delta_{C}/\gamma \sim 5.3$  and $\delta_{SC}/\gamma \sim -2.1$. The inset (b) shows the spring constant and damping as a function of $\delta_{C}/\gamma$, K and $\Gamma$ where the vertical red line represent the stable optical trap from the experimental data.} 
\label{fig:2}
\end{centering}
\end{figure}

\begin{figure}[t!]
\includegraphics[width=\columnwidth]{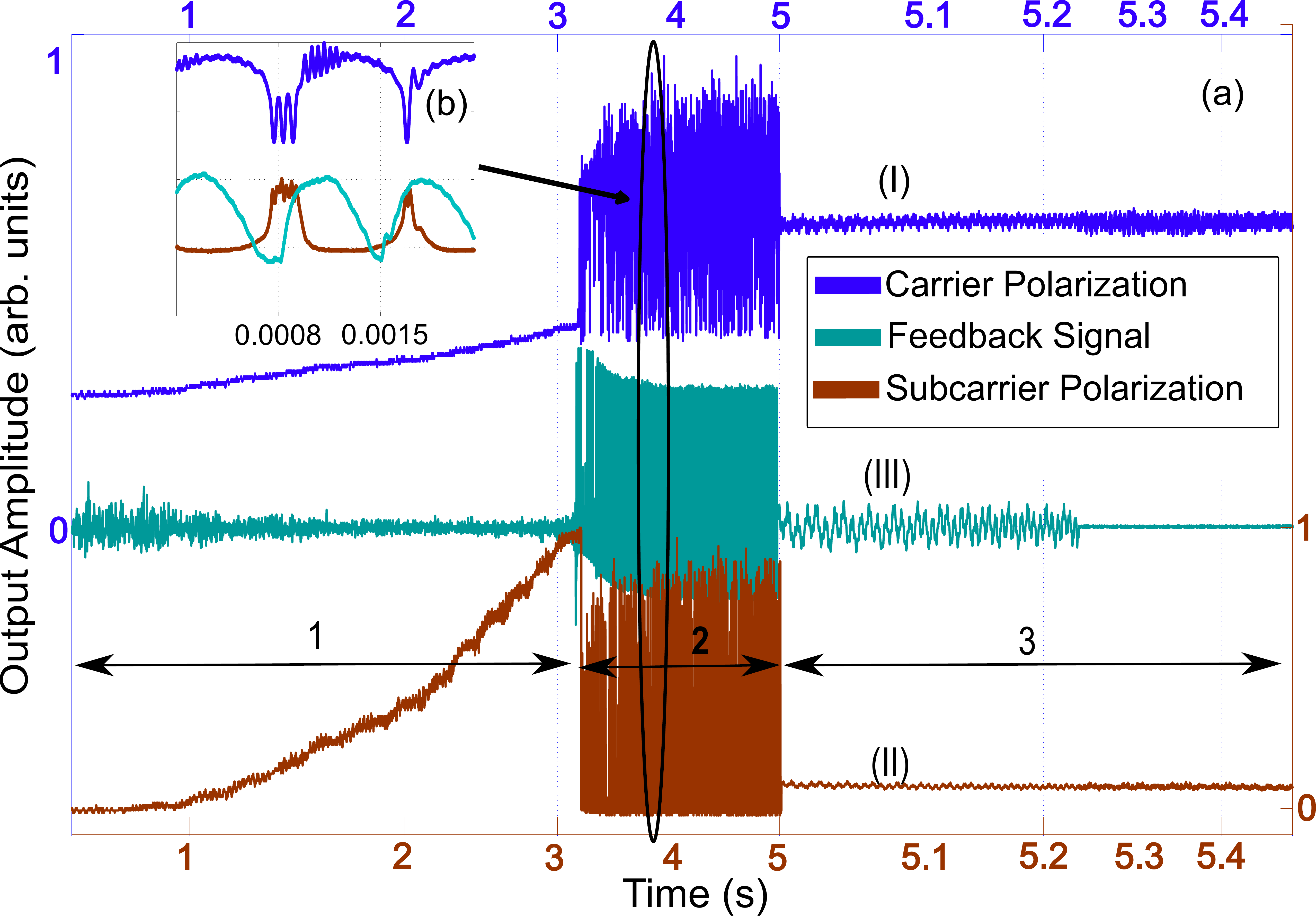}
\caption{The real time sweep data showing the output signal for the C polarization (I), the SC polarization (II), and the feedback to the Amp Mod (III). The region 1 of the plot shows the rise in the amplitude for the C and the SC polarizations, as they scan up the resonant cavity. Oscillations as a result of static instabilities are shown in region 2 of the plot and are magnified in the inset plot (b). The region 3 of the plot shows the system being stable and independent of the feedback control, as shown in region of the plot where the feedback is turned off.}
\label{fig:3}
\end{figure}

Fig \ref{fig:2} depicts the numerical model for operating regimes of our system at a fixed input \agr{coupled power of 42 mW}. The total optical rigidity due to the two polarization field components is plotted as a function of carrier and the sub-carrier detunings. The numerical model is in agreement with our experimentally observed stable optical trap, as can be seen from the locking acquisition of our opto-mechanical system (Fig \ref{fig:3}). The blue dotted line in Fig 2. correspond to the locking acquisition in Fig 3 where the amplitudes for the transmission of the carrier (I), subcarrier (II), and the feedback control signal are shown. The feedback control signal is designed to provide a damping force and is capable of counteracting the optical anti-damping that is dominant during initial locking, which is shown as region 1 in Fig. 2 and 3. When the system enters region 2, the SC crosses onto the other side of resonance, and exerts a strong anti-restoring force. The feedback is unable to counteract an anti-restoring force, and the system oscillates. As the SC detuning increases, and the system moves into region 3, the optomechanical dynamics stabilizes as the restoring force from the C exceeds the anti-restoring force of the SC. At this point, the feedback loop is turned off, and the system remains locked and stable. This does result in slightly higher vibration levels in the absence of the damping feedback loop.



\begin{figure}[h!]
\vspace{0.1in}
\includegraphics[width=\columnwidth]{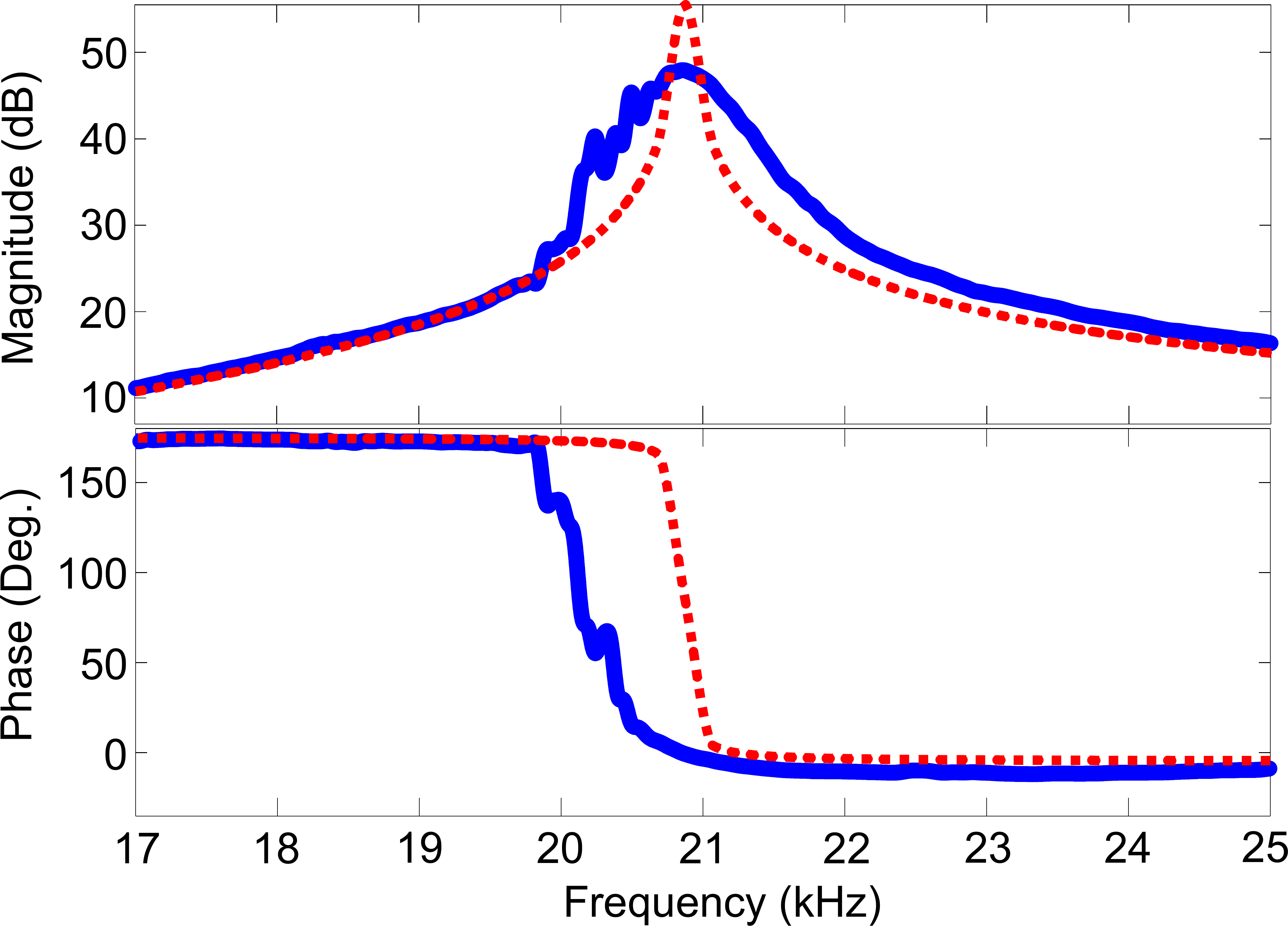}
\caption{The measured transfer function of a signal sent to the amplitude modulator to the transmission photodetector of the carrier is shown (blue), along with the calculated optical response (dotted red). We note that this measurement is performed open-loop, where the feedback signal to the amplitude modulator is turned off here and the cavity is self-stabilized as a result of an optical trap. The disagreement between the measured and calculated Q is attributed to the fact that the the resulting vibrations in the system are sufficient to jitter the intracavity power and modulate the optical spring frequency in the time that it takes to perform this measurement.}
\end{figure}

The inset plot (b) of Fig 2 depicts the \agr{sign of the} total spring constant and the damping coefficient due to the two polarization components, as a function of carrier detuning around the stable optical trap region. The results further correspond to the experimental measurement for the optical spring response at a polarization dependent stable optical trap, discussed above. 

As shown in Fig 4, the mechanical resonance of the microresonator is shifted from 274 Hz to about 21 kHz. The optical trap is stable as can be seen from the decrease in the phase, allowing the system to be operated without any  feedback control. The fluctuations of the optically trapped mirror are relatively large in the performed measurement regime, and there are some non-linearities that are contributing to the noise in this measurement. Fig 4 shows the effects of such fluctuations on the measured transfer function of the oscillator as compared to the calculated transfer function. We have verified that this stabilization is due to polarization and not other effects, such as photothermal effects, by confirming the polarization dependence on the observed stability. We note that by varying the input polarization angle and hence the power in the C and the SC, the observed stability region shifts in agreement with the expected shifts in the detunings of the C and the SC polarizations, and lies in the stable region IV of Fig 2. 

In conclusion, we have demonstrated a polarization dependent stable optical trap for a microresonator based opto-mechanical system, as the outcome of a strong optical spring and optical damping. The dynamics of the system are controlled by radiation pressure and depend on the detunings of the polarization components of the input field. We experimentally demonstrated the stability of the system and confirmed that the deactivation of the feedback control does not render the system unstable. We believe our scheme to be a useful technique for manipulating and stabilizing the dynamics of the vast variety of opto-mechanical systems. 

Due to the simplicity of the technique, the polarization based optical trapping technique has many potential applications in high sensitivity opto-mechanical systems. Since our technique does not depend on absorption, the application can be used without degrading the quantum limited sensitivity of the experiment. 

In the present measurement, the large separation of the two polarizations leads to a smaller than desired optical spring. Thus, we note that it would beneficial to have control over the birefringence effect, so that the difference in the detunings of the carrier and subcarrier, $\delta_C - \delta_{SC}$ could be adjusted, ideally to lie in the range of about $3 \gamma$. \agr{This could be accomplished if both cavity mirrors were made to be birefringent. In that case, one of the mirrors could be rotated with respect to the other, effectively tuning the splitting frequency between the neighboring polarization eigenmodes.} 

This work was supported by the National Science Foundation grant PHY-1150531. This document has been assigned the LIGO document number LIGO-P1600135. \agr{The authors acknowledge D. McClelland of the ANU College of Physics and Mathematical Sciences, for useful discussions and confirmation of similar effects observed in gram scale metal flexures.}

\section{Appendix}

In this supplemental material, we will describe the fabrication details of the microresonator structures used as movable test mass in the stable optical experiment described in the main text.
\section*{Fabrication of Microresonator}

The cantilever microresonators are fabricated from a molecular-beam-epitaxy-grown GaAs/Al$_{\mathrm{x}}$Ga$_{1-\mathrm{x}}$As/In$_y$Ga$_{1-\mathrm{y}}$P heterostructure using a novel double etch stop technique. The material structure consists of a single-crystal distributed Bragg reflector (DBR) based on an epitaxial GaAs (high index, nominal thickness of 77.8 nm for a center wavelength of 1078 nm) / Al$_{0.92}$Ga$_{0.08}$As (low index, thickness of 90.4 nm) multilayer. As the ultimate goal is to operate this structure at cryogenic (liquid $\mathrm{^4}$He) temperatures, the mirror center wavelength is red-shifted to 1078 nm to take into account thermorefractive effects upon cooling \cite{cole08}.

\begin{figure}[h!]
\includegraphics[width=\columnwidth]{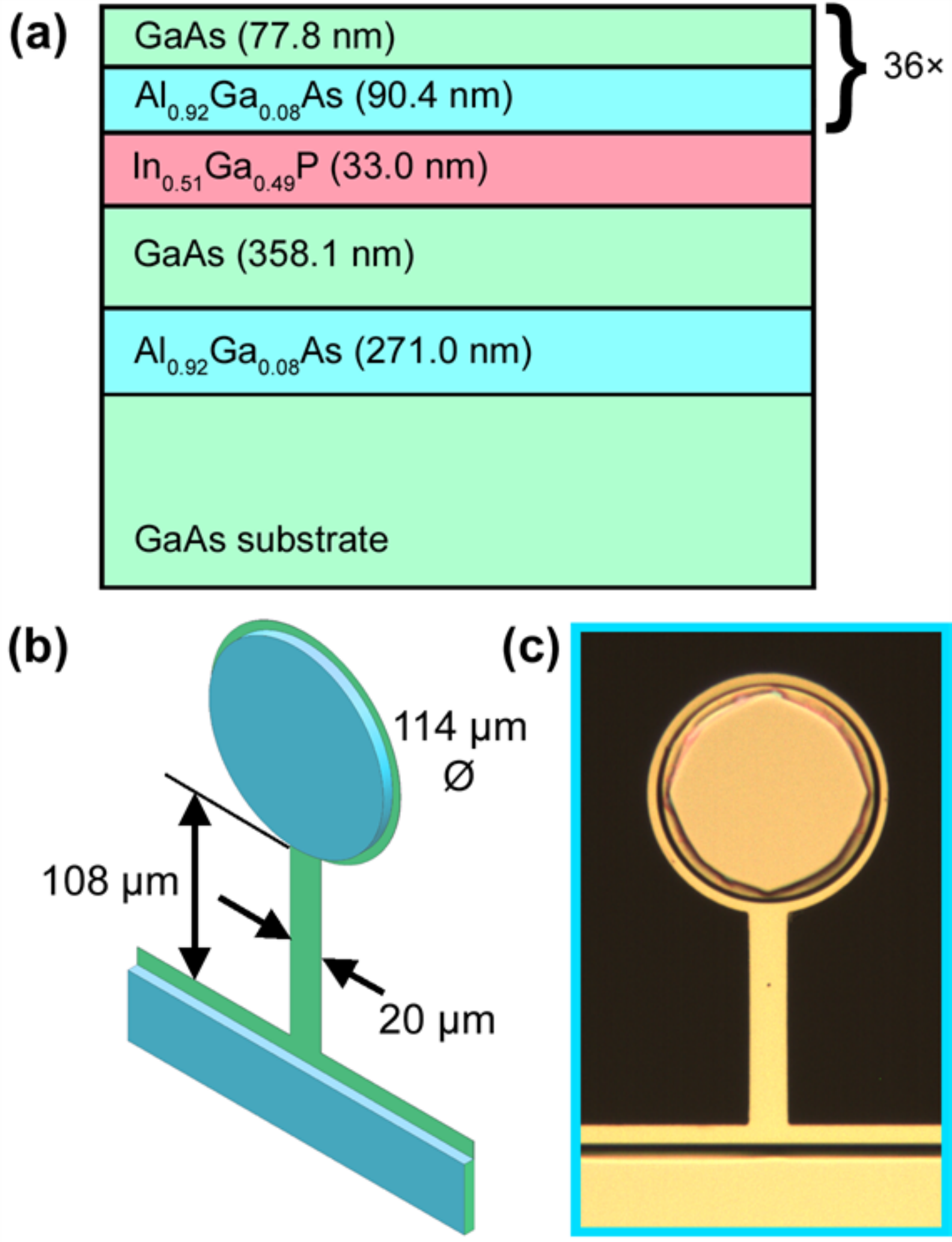}
\caption{Details of the microresonator material structure and mechanical design. {(a)} Cross-sectional schematic of the epitaxial multialyer. From the bottom up, the structure consists of a semi-insulating GaAs substrate, a 271-nm thick Al$_{0.92}$Ga$_{0.08}$AS  backside etch stop layer, a 358.1-nm thick GaAs structural support layer, a lattice-matched $\mathrm{In_{0.51}Ga_{0.49}P}$ etch stop layer, and a 36-period 1064 nm GaAs/$\mathrm{Al_{0.92}Ga_{0.08}As}$ Bragg mirror. The total thickness of the mirror pad, including the topside $\mathrm{In_{0.51}Ga_{0.49}P}$ etch stop is 6.09 \textmu m. (b) Solid model of an example microresonator, consisting of a 108-\textmu m long x 20-\textmu m wide GaAs cantilever, supporting a 114-\textmu m diameter Bragg reflector. (c) Photomicrograph of a completed device with the same nominal dimensions shown in panel b. }
\label{layer}
\end{figure}

The DBR is grown atop a lattice-matched In$_{0.49}$Ga$_{0.51}$P etch stop, a GaAs structural layer, and a second 3/4-wave optical thickness (271 nm) Al$_{0.92}$Ga$_{0.08}$ as etch stop. Unlike the full-thickness mirror structures explored in our earlier work \cite{cole12}, in this design a circular mirror pad is fabricated on a thin film of GaAs in order to de-couple the optical and mechanical properties of the structure, enabling in this case a significant reduction in both the resonator effective mass and spring constant, while maintaining a smaller resonator footprint. This separately optimized structure is similar to previous demonstrations of epitaxial and dielectric resonator designs \cite{sugihwo98,cole05,cole08b,groeblacher09}. \agr{A schematic details of the DBR structure layer and the dimensions of the microresonator are shown in Fig \ref{layer}, below.}

To begin the fabrication procedure, the resonator mirror pad is first defined via optical contact lithographically, by patterning small discs ranging from 114-154 \textmu m in diameter. The resist pads are then used as a mask for vertical etching of the DBR layers using a SiCl$_{4}$-based inductively-coupled plasma etch process. The etch depth is monitored in real time using an in-situ laser interferometer and carried out until only one mirror period remains above the first etch stop. A phosphoric acid based wet etch (H$_3$PO$_4$:H$_2$O$_2$:DI) is then used to remove the final DBR layer pair, stopping with a selectivity of at least 35:1 on the underlying In$_{0.49}$Ga$_{0.51}$P. Minimizing the total time of this wet etching step avoids excessive undercutting of the DBR structure, maintaining the as-designed diameter to the $<$5 \textmu m level.

After definition of the epitaxial mirror pad, the first etch stop is then removed with a dilute HCl solution, stopping with near infinite selectivity on the underlying GaAs structural layer. The addition of a lattice matched ternary In$_{0.49}$Ga$_{0.51}$P film atop the GaAs structural significantly simplifies the microfabrication procedure due to the excellent chemical selectivity between the DBR and etch stop, as well as between the etch stop and underlying GaAs film, allowing excellent control over the mirror pad geometry and structural layer thickness. It is important to note that previous measurements on free-standing strained In$_\mathrm{y}$Ga$_{1-\mathrm{y}}$P resonators has shown the potential for high mechanical quality factors in this material system \cite{cole14}, thus we anticipate no additional mechanical losses from the addition of this layer to the resonator structure. Moreover, the surface of the thin GaAs support layer is only exposed to wet chemical processes and avoids potentially damaging reactive ion etching steps that may lead to excess mechanical losses. These structures ultimately show comparable room temperature quality factors (on the order of 20,000 at 300 K) to our previous low-frequency resonators sculpted from a full-thickness DBR \cite{cole12}. To complete the definition of the lateral geometry of the resonator, a second lithography and ICP etch step is employed. Here, the etch progresses through the GaAs support layer, Al$_{0.92}$Ga$_{0.08}$As backside etch stop layer, and into the GaAs growth substrate, ensuring vertical sidewalls on the final GaAs cantilever structure. The chips are then thinned using a mechanical lapping process to a final thickness of approximately 200 \textmu m (original substrate thickness of 675 \textmu m), re-polished, and, following a thorough clean, the chip is inverted and a silicon nitride (Si$_\mathrm{x}$N$_\mathrm{y}$) hard mask is deposited on the backside of the GaAs growth wafer via plasma enhanced chemical vapor deposition (PECVD). After temporarily mounting the chip to a glass handle using a high temperature wax, a backside lithography step is used to define windows in the PECVD hard mask. The windows defined in this step will ultimately be used to release the micromechanical devices. An SF$_6$-based RIE process is implemented to pattern the Si$_\mathrm{x}$N$_\mathrm{y}$ film and the devices are finally undercut and left free-standing by selectively removing the underlying GaAs growth substrate with a selective H$_2$O$_2$:NH$_4$OH-based wet chemical etch. This process is carried out in an ultrasonic bath to ensure removal of any passivating films formed in etching. This process ultimately terminates on the backside Al$_{0.92}$Ga$_{0.08}$As etch stop layer, while the sidewalls of the cantilever and Bragg mirror are protected by the mounting wax layer. To clean up and free the resonators, the etch stop is removed in a dilute hydrofluoric (HF) acid solution and, after rinsing thoroughly, the samples are soaked in acetone to remove the protective wax and demount the chips from the glass handle. Finally, the samples are transferred to an ethanol bath and a critical point dryer is used to avoid collapse of the free-standing resonators during the solvent removal stage.

\bibliographystyle{ieeetr}
\bibliography{biblioDOS}

\end{document}